\documentclass[11pt,a4paper]{article}
\usepackage{graphicx}
\usepackage{jheppub}
\usepackage{subfig}
\usepackage[latin1]{inputenc}
\usepackage{amsmath}
\usepackage{amsfonts}
\usepackage{amssymb}
\usepackage{setspace}
\usepackage{mathtools}
\usepackage{pstricks}
\usepackage{color}

\newcommand{\ba}{\begin{eqnarray}}
\newcommand{\ea}{\end{eqnarray}}

\newcommand{\ms}{m_\sigma}

\newcommand{\sect}[1]{\mbox{section\ \ref{#1}}}
\newcommand{\Sect}[1]{\mbox{Section\ \ref{#1}}}

\title{Thermal blocking of preheating}

\author[a]{Rose Lerner,}
\author[b]{Anders Tranberg}

\affiliation[a]{Deutsches Elektronen-Synchrotron DESY, Notke{\ss}trasse 85, 22607 Hamburg, Germany} 
\affiliation[b]{Faculty of Science and Technology, University of Stavanger,
N-4036 Stavanger, Norway} 

\emailAdd{rose.lerner@desy.de}
\emailAdd{anders.tranberg@uis.no}

\abstract{
The parametric resonance responsible for preheating after inflation will end when self-interactions of the resonating field and interactions of this field with secondary degrees of freedom become important. In many cases, the effect may be quantified in terms of an effective mass and the resulting shifting out of the spectrum of the strongest resonance band.
In certain curvaton models, such thermal blocking can even occur before preheating has begun, delaying or even preventing the decay of the curvaton.
We investigate numerically to what extent this thermal blocking is realised in a specific scenario, and whether the effective mass is well approximated by the perturbative leading order thermal mass. We find that the qualitative behaviour is well reproduced in this approximation, and that the end of preheating can be confidently estimated.}

\preprint{DESY 15-019}


\begin{document}
\maketitle

\section{Introduction}
\label{sec:introduction}

At the end of inflation, the Universe was reheated through the transfer of energy from the inflaton field to a number of coupled scalar, vector and fermion fields. This occurred through a combination of perturbative decay and non-perturbative preheating. Primary particles created in such processes subsequently decayed into other secondary degrees of freedom, and the whole system ultimately thermalised to produce a radiation-dominated Universe.

Resonant preheating occurs because the post-inflationary oscillations of the inflaton are in resonance with certain momentum modes of the primary field(s) \cite{preheating}. The final particle number depends on the strength and duration of the resonance. In general, there is an infinite tower of resonance bands, but the strength of the resonance drops significantly beyond the first band. In the simplest case, a first-band resonance occurs if there is a momentum mode ${\bf k}$ satisfying
\ba
\label{eq:firstband}
{\bf k}^2+M_{\rm eff}^2=m_{\rm inf}^2,
\ea
where $m_{\rm inf}$ is the frequency of inflaton oscillation. $M_{\rm eff}$ is the effective mass of the field coupled to the inflaton; it changes in time due to both interactions with other fields and self-interactions. This paper deals with the blocking of the resonance due to the background temperature causing an effective thermal mass to be generated (termed "thermal blocking"). This could happen either as the system heats up, such that a resonance that is effective at the beginning of the preheating process shuts off when the temperature (or out-of-equilibrium spectrum) is established, or due to an already-present thermal background, such as in the curvaton model. We do not discuss in this paper the backreaction of the resonant field on the oscillating field, which modifies the oscillation frequency and can also end preheating.

The same physical process can occur in curvaton models \cite{curvaton}, where a scalar field distinct from the inflaton is present, but has negligible energy density during inflation and decays long after the end of inflation. Before reheating, the curvaton comes to almost dominate energy density of the Universe, and upon reheating, its isocurvature perturbations become the adiabatic curvature perturbations observed in the CMB. Similarly to the inflaton case, the curvaton can decay either perturbatively or through non-perturbative preheating. In contrast to the inflaton case, there generically exists a thermal background at the start of curvaton decay, formed by the inflaton's decay products. The process of curvaton reheating has been the subject of recent attention \cite{curvatondecay}, particularly the report that preheating in the curvaton model into Higgs particles can be substantially delayed by the presence of the thermal background \cite{dani}. This result is an integral component of the minimal curvaton-higgs (MCH) model, in which the curvaton is coupled directly to the Standard Model Higgs boson \cite{mch}. Without this blocking, the model would become unviable, unless the Higgs-curvaton coupling were tuned to be tiny, because reheating would occur while the curvaton was too subdominant. Thus, the viability of the MCH model depends on whether the analytical calculations in \cite{dani} hold up beyond the simple LO perturbative estimate of the thermal mass. 

The onset and end of non-perturbative instabilities has been studied numerically for some time in the context of resonant preheating (see for instance \cite{respre_num}), tachyonic preheating (see \cite{tacpre_num}), self-resonance (see \cite{Hertzberg}), and also for heavy-ion collisions (see \cite{QCD_num}). Because the particle numbers are often very large, a classical-statistical approach is usually adopted, but also quantum dynamics has been studied in this context \cite{2PI,clasand2PI}. In most cases, the main emphasis has been on the kinetic thermalisation phase and the back-reaction on the field driving the instabilities, rather than in terms of a thermal mass. For a general review of nonperturbative reheating, see \cite{Amin_review}.

The two aims of this paper are to determine the conditions under which the thermal blocking effect is present; and to determine the extent to which thermal blocking is well described by an effective leading order thermal mass. 

Our paper is organised as follows. In \sect{sec:blocking} we introduce our model and our analysis of preheating, including the concept of the thermal mass in the leading order (LO) approximation. We also discuss the timescale of the various physical processes. In \Sect{sec:numerics} we present our numerical results for thermal blocking, showing both the results in the LO approximation and once dynamical secondary fields are introduced. We conclude in \sect{sec:conclusion}.

\section{Thermal blocking of preheating}
\label{sec:blocking}

Our numerical investigation uses the curvaton model as an example. This is because the thermal background is produced well before curvaton preheating, by the decay of the inflaton. Thus, it is much easier to model than a pure inflaton model, where the thermal background is produced {\em during} preheating.

\subsection{Model}
\label{sec:model}

We consider a very simplified model of the Standard Model Higgs field $\phi$ coupled to both the curvaton scalar $\sigma$ and $N_f$ additional light degrees of freedom $\xi_i$, $i=1,....,N_f$. These light fields represent the fermions and gauge fields of the Standard Model, in terms of decay channels and thermal back-reaction on the Higgs field. We take them to be massless.
 
We will assume that the Standard Model degrees of freedom ($\phi$ and the $\xi_i$) are initially in thermal equilibrium at a temperature $T$, and that $\sigma$ is homogeneous and oscillating. The field $\sigma$ is coupled only to the Standard Model Higgs. Due to this coupling, there is potential for resonant production of Higgs excitations. The additional fields $\xi_i$ act as secondary fields; they are not coupled to the oscillating field $\sigma$. We neglect the expansion of the Universe. 

The Standard Model Higgs field is a 2-component complex scalar field given by
\ba
\phi = \frac{1}{\sqrt{2}}\left(
\begin{array}{c}
\phi_1+i\phi_2\\\phi_3+i\phi_4\end{array}\right),
\ea
and thus the action is
\ba
S= -\!\int d^4x \! \left[
\partial_\mu\phi^\dagger\partial^\mu\phi-\mu^2\phi^\dagger\phi+\lambda(\phi^\dagger\phi)^2+g^2\sigma^2\phi^\dagger\phi+h^2\phi^\dagger\phi\sum_i\!\xi_i^2+\frac{1}{2}\!\left(\partial_\mu\xi_i\partial^\mu\xi_i\right)
\right] 
\ea
with the known constants
\ba
\mu^2 = \frac{m_H^2}{2} = \frac{126^2\textrm{ GeV}^2}{2},\qquad
\frac{\mu^2}{\lambda}=v^2=246^2\textrm{ GeV}^2,
\ea
and an unknown coupling $g$. The function $\sigma^2=\sigma^2(t)$ is a priori unknown; for simplicity we take it to be
\ba
\sigma^2(t) = \sigma_0^2 \cos^2 \left(m_\sigma t\right),
\ea
with some new unknown constants $\sigma_0$, and $m_\sigma$. Note that, in a fully dynamical setup, the field $\sigma$ would also be dynamical and interact with its environment (we do not implement this here).

The equation of motion for the Higgs field is then
\ba
\left[\partial_\mu\partial^\mu-\mu^2+2\lambda\,\phi^\dagger\phi+g^2\sigma^2(t)+h^2\sum_i\xi_i^2\right]\phi (x,t) =0,
\ea
and for the secondary fields,
\ba
\left[\partial_\mu\partial^\mu+2h^2\phi^\dagger\phi\right]\xi_i (x,t) =0.
\ea
Given $g$, $\sigma_0$, $m_\sigma$ and $h$, these can be solved numerically. We have chosen the secondary fields to be massless, which is approximately true for reheating temperatures beyond the electroweak scale (except for the top quark). 
Because the top Yukawa coupling is of order one, and the electroweak gauge coupling is not much smaller than unity, the effective Higgs-SM coupling $h$ should also be taken to be of order one.

\subsection{Resonant preheating}
\label{sec:res}

In our model, both primary $\phi_{1,..,4}$ and secondary fields $\xi_{1,..,N_f}$  start out in thermal equilibrium. Thus, the Fourier components of each field and its conjugate momentum $\pi_i= \partial_t\phi$ (in continuum notation), which are given by
\ba
\phi_i = \int \frac{d^3k}{(2\pi)^3}\phi_i({\bf k}) e^{i{\bf k x}},\qquad\pi_i = \int \frac{d^3k}{(2\pi)^3}\pi_i({\bf k}) e^{i{\bf k x}},
\ea
obey\footnote{We do not include the ``quantum half'' (zero point energy). Doing so would require renormalisation of the mass, but otherwise results in qualitatively similar results.}
\ba
\label{eq:init}
\langle\phi_i^\dagger({\bf k})\phi_i({\bf k})\rangle=\frac{n_{\bf k}}{\omega_{\bf k}},\quad \langle\pi_i^\dagger({\bf k})\pi_i({\bf k})\rangle=n_{\bf k}\omega_{\bf k},
\ea
with $\omega_{\bf k}^2=m_{\rm initial}^2+{\bf |k|}^2$. Particle numbers are given by the Bose-Einstein distribution
\ba
\label{eq:initT}
n_{\bf k}(T) = \frac{1}{e^{\omega_{\bf k}/T}-1}.
\ea
Similar expressions hold for the $\xi_i$ fields. 

\noindent The equation of motion for the Higgs field in momentum space is (including only local self-energy components; see section \ref{sec:thermalmass} for details)
\ba
\label{eq:eomp}
\left[\partial_t^2+{\bf |k|}^2-\mu^2+\lambda\langle\phi^\dagger\phi\rangle+M^2(T)+g^2\sigma_0^2\cos^2(m_\sigma t)\right]\phi_{\bf k}=0.
\ea
A good choice for the initial state mass of the Higgs field is 
\ba
m_{\rm initial}^2=-\mu^2+M^2(T)+g^2\sigma_0^2,
\ea
with the assumption that this quantity is positive. As stated previously, $\xi_i$ are massless. The possibility of resonant preheating can best be seen by rewriting (\ref{eq:eomp}) into the Mathieu equation using
\ba
A_{\bf k}=\frac{{\bf |k|}^2-\mu^2+\lambda\langle\phi^\dagger\phi\rangle+M^2(t)+g^2\sigma_0^2/2}{m_\sigma^2},\qquad q=-\frac{g^2\sigma_0^2}{4m_\sigma^2}, \qquad \tau = m_\sigma t,
\ea
to get 
\ba
\partial_\tau^2\phi_{\bf k}+\left(A_{\bf k}-2q\cos(2\tau)\right)\phi_{\bf k}=0.
\ea
This equation has resonant solutions around $A_{\bf k}=1,4,9,...$, corresponding to 
\ba
\frac{{\bf |k|}^2}{m_\sigma^2} = \frac{A_{\bf k}+\mu^2-\lambda\langle\phi^\dagger\phi\rangle-M^2(t)-g^2\sigma_0^2/2}{m_\sigma^2}.
\ea
In the following, we take the fiducial values $g^2=0.1$, $\sigma_0/m_\sigma=1$, and  $m_H/m_\sigma=0.3$. Note, that because $\sigma$ is not a dynamical field, only the combination $g^2\sigma_0^2/m_\sigma^2=0.1$ is relevant. For these values, in the first resonance band we find
\ba
\label{eq:shift}
\frac{{\bf |k|}^2}{m_\sigma^2}=0.995-\lambda\langle\phi^\dagger\phi\rangle-M^2(t).
\ea
The growth index of the first resonance band is thus
\ba
\mu_{\rm res}\simeq q/2=\frac{g^2\sigma_0^2}{8m_\sigma^2}=0.0125,
\ea
and the following resonance bands are weaker.

According to (\ref{eq:shift}), when $ M^2(t)\simeq 1$ the first resonance band becomes inactive, and the preheating process ceases to be effective. We demonstrate this numerically in \Sect{sec:numerics} in two ways; in the first we insert $M^2(t)$ (\ref{eq:LOres}) by hand; and in the second we simulate $N_f$ scalar fields directly as dynamical degrees of freedom. This allows us to test the analytical calculation of thermal blocking.

\subsection{Thermal and out-of-equilibrium masses} 
\label{sec:thermalmass}

The effective mass of the primary particles to leading order in a coupling expansion, is
\ba
\label{eq:Meff}
M_{\rm eff}^2(t)=-\mu^2+g^2\sigma^2(t)+M^2_{\rm self}(t)+M^2_{\xi}(t).
\ea
To this order, the self-interaction from the quartic Higgs coupling is 
\ba
\label{eq:selfint}
M^2_{\rm self}(t)=\lambda \langle\phi^\dagger\phi\rangle,
\ea
with the correlator computed in the out-of-equilibrium state given by the resonance spectrum of the preheated Higgs field. $M_{\rm self}$ is therefore time-dependent, and may also have a thermal component. The contribution to the effective mass from interactions between primary and secondary fields is, to leading order and in equilibrium at temperature $T$, 
\ba
\label{eq:LOres}
M^2_{\xi}(t)=h^2 \langle\xi^2\rangle=\frac{h^2N_fT^2}{12},
\ea
where the last equality applies for massless fields in thermal equilibrium\footnote{We note that at zero temperature this is probably not a good approximation, since the $\xi_i$ will have a mass proportional to the Higgs vev, $m_\xi^2 = \frac{1}{2}h^2 v^2$. At finite temperature, there will also be thermal masses for the $\xi_i$ as well, which at LO and in the high-temperature limit are given by $M^2_{\xi}=h^2 \langle\phi^\dagger\phi\rangle\simeq\frac{h^2T^2}{3}$.
}.
Due to $M_{\rm self}$ and $M_{\xi}$, there can therefore be thermal blocking of the resonance in the primary field. This occurs if either the temperature $T$ or coupling $N_fh^2$ are large enough such that the sum of mass components (\ref{eq:Meff}) becomes larger than the driving frequency $m_\sigma$.

The total effect of interactions with the secondary fields (and self-interactions) is not always well described by (\ref{eq:Meff}) (this is the expression used in \cite{dani} for analytical calculations). In general, one should instead use the complex, non-local selfenergy $\Sigma$, which appears in the corresponding evolution equation in momentum space as
\ba
\left[\partial_t^2+{\bf |k|}^2-\mu^2+g^2\sigma^2(t)\right]\phi_{\bf k}(t)+\int_{\bf k'}\int_{t'}\Sigma({\bf k, k'},t,t')\phi_{\bf k'}(t')=0.
\ea
Under certain conditions, $\textrm{Re}[\Sigma]$ can be thought of as a thermal mass contribution and $\textrm{Im}[\Sigma]$ as decay or a damping rate \cite{local_approximation}. Such an identification requires that the self-energy is approximately local in time. At LO in a loop expansion, $\Sigma$ is real, local and equivalent to the expression used in the analytical calculations (\ref{eq:LOres}). However, the complete self-energy requires more complicated methods to compute, which we implement numerically in a classical (as opposed to fully quantum) theory. This classical-statistical approximation is known to be very good for high temperatures and large particle numbers \cite{classical,clasand2PI}.

 \subsection{Time scales}
 \label{sec:timescales}
 
We may identify a number of time-scales in the problem. The basic time-scale is the oscillation of the driving field $\sigma$, which is $\tau_{\rm driving}\simeq m_\sigma^{-1}$. The resonance builds up in a time of the order $\tau_{\rm resonance} = (10-100)\, \tau_{\rm driving}$. The primary particles produced by the resonance then decay. If this decay is into other Higgs modes, then the timescale is given by the perturbative decay width of the Higgs at finite temperature, giving $\tau_{\rm decay, 1}\propto \Gamma_{\phi}^{-1}\propto \lambda^{-2}m_H^{-1}$. If the decay is into excitations of the secondary fields $\xi$, the timescale is given by another perturbative decay width, $\tau_{\rm decay, 2}\propto \Gamma_{\xi}^{-1}\propto h^{-4} m_H^{-1}$. These secondary particles then thermalise, and the time-scale for kinetic thermalisation (redistribution of the particles into a semi-thermal spectrum) is of order a hundred times the decay time $\tau_{\rm kin, 1/2}\simeq 100 \, \tau_{\rm decay, 1/2}$, and the timescale for complete chemical equilibrium is one or two orders of magnitude larger than that, $\tau_{\rm chem, 1/2}\simeq (10-100)\, \tau_{\rm kin, 1/2}$. In addition, in the case of an expanding Universe, there is the Hubble time, $\tau_{H}=H^{-1}$.
 
We can find the ordering of the timescales using a few facts: a succesful resonance requires $m_\sigma>m_H$; $h^2=\mathcal{O}(1)$ (Standard Model couplings); and $\lambda = m_H^2/(2v^2)=0.13$. We therefore expect that $\tau_{\rm decay, 2 }\lesssim\tau_{\rm decay, 1}\ll \tau_{\rm kin, 1/2 }\ll\tau_{\rm chem, 1/2}$.  We also assume that the Hubble expansion is negligible, an assumption that depends strongly on what happens in the inflaton sector\footnote{If the energy is still in the inflaton, the expansion rate could be substantial. If it has been dumped into the thermal background of secondary and Higgs fields, they will either be at 100 GeV temperatures, in which case expansion is negligible; or they will have been reheated to very large temperatures indeed, in which case all mass-scales in the above time-scale estimates must be replaced by that temperature.}. The simulations performed here end before kinetic equilibrium has completed, so before $\tau_{\rm kin, 1/2 }$. 
 
In the case where $m_\sigma\gg m_H$  and $m_\sigma\gg T$, the resonance completes much faster than any other processes can react. It is then not obvious whether the microscopic processes giving rise to the thermal masses can ``keep up''. However, the resonance bands are then high above the scales $m_H$ and $T$ anyway, and we would expect no thermal blocking, except for a small shift in the peak location. We do not discuss this case further.
 
However, if $m_\sigma >m_H$ and  $m_\sigma> T$ (but not $\gg$), $\tau_{\rm resonance}\simeq \tau_{\rm decay, 1/2}$, and thermal blocking may kick in, if the details of the dynamics generate a mass, and if the excitations do not decay away too fast. It is this regime that we investigate in this paper. It is also the regime where it is likely that the LO result (\ref{eq:LOres}) is incomplete.
 
\subsection{Numerical implementation}
\label{sec:numerics}

We discretize the system on a $64^3$-site spatial lattice and solve the classical equations of motion in real-time with a time-step of $dt=0.05$. All dimensionful quantities will be quoted in units of $m_\sigma$, which in lattice units is unity, $am_\sigma=1$. This means that the basic frequency is well inside the available dynamical range of the lattice, and since we are mostly interested in IR physics, the rather large lattice spacing will not be crucial. We generate $N=64$ random classical realisations\footnote{In addition to summing over random realisations, we sum over the four Higgs degrees of freedom, and over hyper cubic symmetry, i.e.\ identifying modes of momentum ${\bf k}$ that only differ by permutations of $\pm k_x,\pm k_y,\pm k_z$. This gives up to 48 different combinations that are averaged over. } of the $\phi$ and $\xi$ fields for each set of parameters $h$, $T$, $N_f$. We run the simulations until $t=1000\,m_\sigma^{-1}$, which is easily enough to have a strong resonance, but shorter than the thermalisation time of the system. Hence we expect to see a clear peak in the spectrum, although decays will also have an effect, because the time-scales involved are around $\tau_{\rm decay, 1/2}$

The instantaneous particle number distribution can then be computed from the inversion of (\ref{eq:init})
\ba
n_{\bf k} =\left(\langle\phi_i^\dagger({\bf k})\phi_i({\bf k})\rangle\langle\pi_i^\dagger({\bf k})\pi_i({\bf k})\rangle\right)^{1/2}.
\ea

\section{Numerical Results}
\label{sec:numerics}

In the following, we present a number of different approximations to the effective mass. We begin by showing the effect of adding a ``by hand'' thermal mass (\sect{sec:LOthermal}). We then compare this to the case where the secondary fields are dynamical, and show how the results depend on the number of fields involved (\sect{sec:dynamical}).

\subsection{The LO thermal mass by hand}
\label{sec:LOthermal}

\begin{figure}
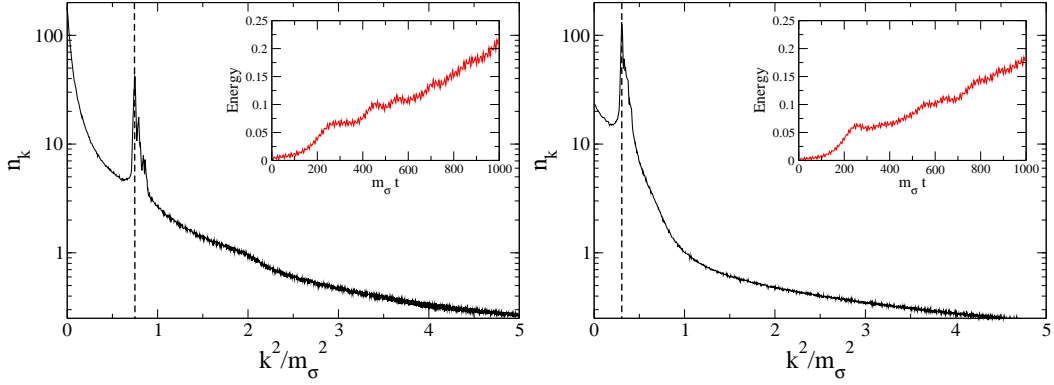

\begin{center}
\epsfig{file=figs/Spectrum_example.eps,width=0.45\textwidth}
\epsfig{file=figs/Spectrum_example2.eps,width=0.45\textwidth}
\caption{\label{fig:example1}The particle number after preheating at $t=1000\,m_\sigma^{-1}$, corresponding to approximately 160 inflaton oscillations. Notice the logarithmic scale. Inset is the energy in the preheated field(s). The Higgs field is self-interacting and coupled to the ``by-hand'' inflaton, but has no coupling to any other fields. Shown without an additional mass (left), and with a mass of $M^2=0.5m_\sigma^2$ (right). Note how the additional mass shifts the resonance peak to smaller ${\bf |k|}^2$.}
\end{center}
\end{figure}

Fig.~\ref{fig:example1} (left) shows the particle number distribution as a function of ${\bf |k|}^2$ at time $t=1000\,m_\sigma^{-1} $ when the interaction with the $\xi_i$ fields is turned off, but the self-interaction is still present. Note that the x-axis is the momentum squared, and the y-axis is logarithmic. We have taken a reference temperature $(T/m_\sigma)^2=2$. We see a clear primary peak at ${\bf |k|}^2/\ms^2=0.75$ (with some additional structure), on the background of the initial thermal spectrum. As expected, it is close but not exactly at the first resonance ${\bf |k|}^2/\ms^2\simeq 1$, because the self-interaction provides an effective mass. There are also secondary peaks due to the self-interactions, but it is not possible to see the second resonance band around ${\bf |k|}^2/\ms^2\simeq 4$. The inset figure shows the energy in the Higgs field, which rises monotonically, athough not exactly linearly with time. The net transferred energy is not significantly affected by the shift in the peak. 

In Fig.\ref{fig:example1} (right) we show the spectrum when we have added by hand a ``thermal mass'' $M^2(T)=0.5$. The peak is now shifted to $k^2/\ms^2=0.31$, which, within the width of the peak, is fairly consistent with a shift of $0.5$. We still do not have any dynamical secondary fields $\xi$ present; their effect is mimicked by the by-hand thermal mass. 

\begin{figure}
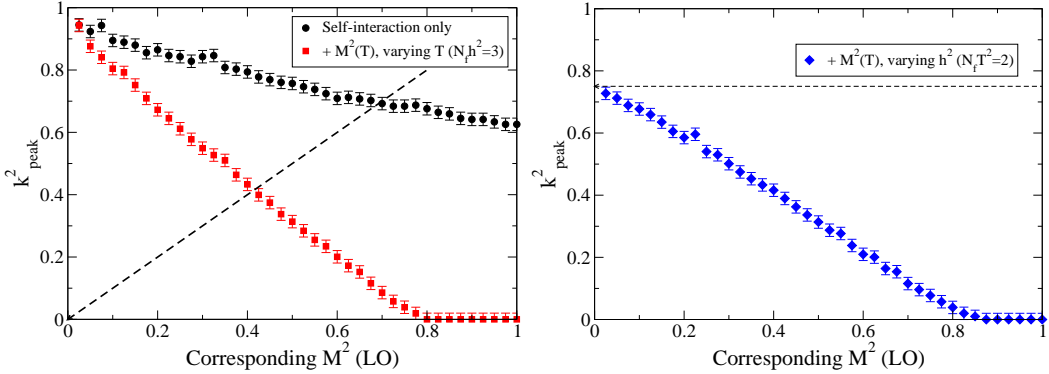

\begin{center}
\epsfig{file=figs/Thdep1_example3.eps,width=0.45\textwidth}
\epsfig{file=figs/Thdep2_example3.eps,width=0.45\textwidth}
\caption{\label{fig:byhand}The position of the resonance peaks as a function of the corresponding effective (LO) mass, as a function of temperature $T$ (left) and coupling $h$ (right). {Left:} Black circles (top) show only the effect of Higgs self-interaction; red squares (bottom) also includes the LO mass; dashed line is explained in the text. {Right:} The dashed line here is the result including only self interaction; points include the LO mass, with fixed $T^2=2$ for the Higgs field.}
\end{center}

\end{figure}

Fig.~\ref{fig:byhand} (left) shows the peak location as a function of the ``by-hand'' $M^2(T)$ of the system. To obtain the figure, we repeated the calculation used to make Fig.~\ref{fig:example1} (left) with varying $M^2(T)$. The error bars give an estimate of the systematic uncertainty due to the width of the resonance peaks in the numerics (compare to Fig.~\ref{fig:example1}). 
The circular black points (top) are equivalent to Fig.~\ref{fig:example1} (left), i.e.\ include only self-interactions, calculated at various initial temperatures of the Higgs field. As expected, just changing this initial temperature changes the location of the peak.
The square red points (bottom) are equivalent to Fig.\ref{fig:example1} (right) i.e.\ include a by-hand thermal mass, in addition to the self-interaction. To obtain these points, we again varied the initial temperature, but in addition changed $M^2(T)$ by varying $T$, while keeping $N_fh^2$, in accordance with (\ref{eq:LOres}). 
Thus, the shift represented by the square red points is the sum of the effect of the explicit $M^2(T)$, and the self-interaction of the Higgs, the circular black points. 

The dashed line in the figure shows $M^2(T) = {\bf |k|}^2$. If the effects of thermal masses were strictly additive (i.e.\ as assumed in an analytical calculation), then the position of the peak should reach ${\bf |k|}^2=0$ at the value of $M^2(T)$ where the dashed line crosses the circular black points, i.e.\ $M^2(T) = 0.7$. We see that in fact the peak does not leave the spectrum until slightly larger $M^2(T)$, 0.8 instead of 0.7. Thus, the thermal masses are not strictly additive.

In Fig.~\ref{fig:byhand} (right), we show the location of the peak as a function of $M^2(T)$, when we imagine that the variation is due to a change in $h^2$, with $N_fT^2$ kept fixed. In this case, the contribution from the self-interaction is constant (symbolised by the dashed line; see Fig.~\ref{fig:example1}), and the effect of the explicit $M^2(T)$ adds to this contribution. Again, with strictly additive behaviour, one would expect that the peak leaves the spectrum at $M^2(T)=0.75$. There is however a small delay to this, and the peak leaves the spectrum at around $M^2(T)=0.83$.

In this section, we have shown that a large-enough by-hand thermal mass (\ref{eq:LOres}) causes the resonance to move out of the available modes (mass becomes too big), and preheating to stop. The behaviour is convincingly linear, and different effects approximately (but not exactly) add up. Higher resonance modes do not seem to provide a sufficiently fast decay to prevent this thermal blocking.

\subsection{Dynamical secondary fields}
\label{sec:dynamical}

\begin{figure}
\begin{center}
\epsfig{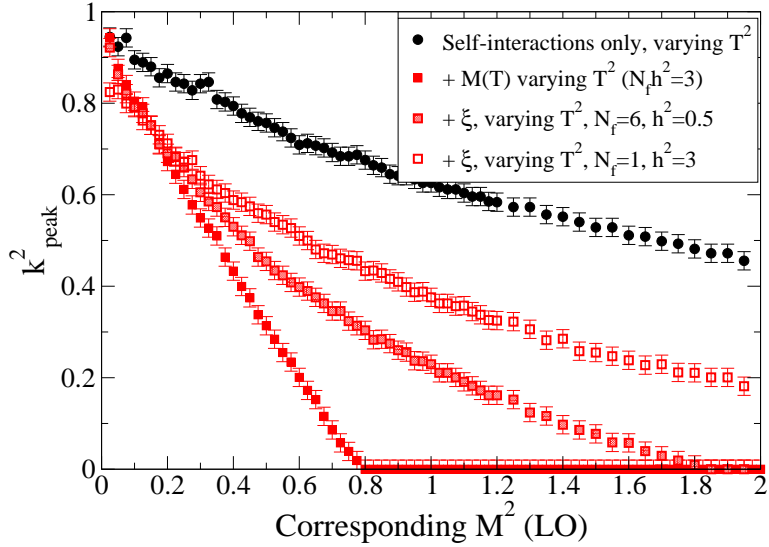}
\caption{\label{fig:fullT}Shows the position of the resonance peak when varying $T$, against the corresponding effective (LO) mass (\ref{eq:LOres}). Filled symbols: Self-interactions only (circles; top) and self-interactions plus LO effective mass (squares; bottom). Open symbols (second from top): Full dynamical light fields with $N_f = 1$.  Shaded symbols: Full dynamical light fields with $N_f = 6$. }
\label{fig:fullT}
\end{center}

\end{figure}

We now consider the full field dynamics, beyond LO, and investigate whether the resonance peak is still shifted out of the spectrum, i.e.\ we determine to what extent the thermal mass is well represented by the LO result. In order to compare to the LO result, Fig.~\ref{fig:fullT} shows again the self-interacting (circular black), and LO (filled red squares) results, copied directly from Fig.~\ref{fig:byhand} (left). These were the results at LO obtained by varying T. In addition, we show two cases with dynamical fields $\xi_i$. The first (open symbols; second from top) has $N_f=1$ and a coupling of $h^2=3$. The second (half-shaded symbols; second from bottom) has $N_f=6$ and $h^2=0.5$. At LO, these results would be equivalent, because $N_fh^2=3$ in both cases. However, beyond LO the degeneracy is lifted. We expect that the large $N_f$, small coupling $h^2$ case should be closer to the LO result, since the expression is precisely leading order in a coupling and/or $1/N_f$ expansion. 

These expectations are confirmed by the simulations. Thermal blocking is still effective, but somewhat less so for the fully dynamical system. By the time the resonance is fully blocked with the LO mass, the peak in the full dynamics is still at $k^2=0.4$ ($N_f=1$) or $k^2=0.3$ ($N_f=6$). We also observe that the larger $N_f$ case is indeed closer to the LO result.

Thermal blocking is complete at $N_f=6$ for $T^2\simeq 7.2$, whereas for $N_f=1$, simple extrapolation suggests a $T^2$ of between twice and three times that is required. We dare not simulate to larger temperatures, for fear of populating modes near the lattice cutoff. 

\begin{figure}
\begin{center}
\epsfig{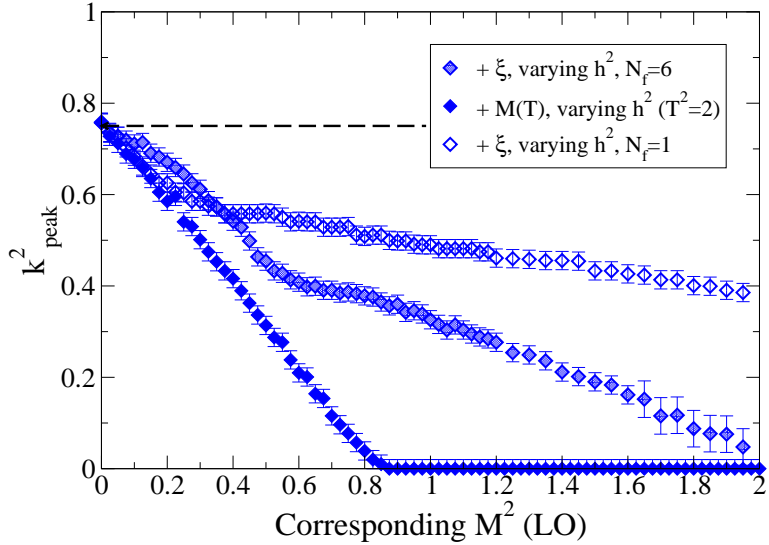}
\caption{\label{fig:fullh}Shows the position of the resonance peak when varying $h$, as a function of the corresponding effective (LO) mass (again through (\ref{eq:LOres})). Dashed line: Self-interactions only. Filled symbols (bottom): Self-interactions plus effective mass. Open symbols (top): Full dynamical light fields with $N_f = 1$.  Shaded symbols (middle): Full dynamical light fields with $N_f = 6$.}
\end{center}

\end{figure}

We now investigate how the results depend on $h$, showing the results in Fig.~\ref{fig:fullh}. For comparison, we show the results including a by-hand thermal mass, keeping $T^2=2$ fixed but varying $h$ (filled blue diamonds; equivalent to the points in Fig.~\ref{fig:byhand}, right). In analogy with Fig.~\ref{fig:fullT}, we show the corresponding curves including the effects of dynamical fields. Open symbols denote $N_f=1$ and half-shaded points $N_f=6$. In this case, we see that the three curves agree well for small couplings $h^2\lesssim 2$ ($M^2(T)<0.3$). The dynamical result with $N_f = 1$ bends off at larger couplings, making the LO thermal mass a less robust approximation in that case. Even for $N_f=6$, the agreement becomes less convincing around $h^2=4$ ($M^2(T)<0.6$).

For $N_f=6$, it is clear that thermal blocking will complete around $M^2(T)\simeq 2$, a factor of between two and three larger than the LO expectation. For $N_f=1$, the picture is less clear. However, a linear extrapolation suggests complete blocking around $M^2(T)=5$, safely within a factor of ten of the LO estimate. 

\section{Conclusion}
\label{sec:conclusion}

Resonant preheating occurs when the frequency of the driving field ($\sigma$) corresponds to the frequency of some mode of the resonating field ($\phi$). In many cases, only the first resonance mode is effectively resonating. If the effective mass of the resonant field is higher than the driving frequency, then there is no preheating because that mode does not exist in the spectrum. Thus, the formation of an effective mass is a mechanism that may end or prevent preheating. The effective mass can be generated through self-interactions or through thermal effects, if the resonant field is coupled to thermalised (or even just populated) fields.

One may expect that this thermal blocking can be modelled by computing the thermal mass to leading order in perturbation theory, and that higher order terms contribute only small corrections. In this paper we investigated this assumption and found that naively using the LO thermal mass is a fair approximation when the number of secondary fields is large ($>6$). In this case, thermal blocking prevents preheating as the resonating mode is shifted out of the spectrum, as represented by
\ba
k_{\rm peak}^2\propto k_{\rm peak}^2(T=0)-cN_fh^2T^2.
\ea
The (model-dependent) constant is $c=1/12$ at LO, but is somewhat smaller in the exact case (by a factor of 1/2 to 1/10; smaller with increasing $N_f$). However, when the number of secondary fields is small ($N_f=1,2$) the LO approximation is poor, and even for large couplings and temperatures, the thermal shifting is ineffective and does not block the resonance.

For models where the primary field is the Higgs, such as the MCH model, the number of secondary degrees of freedom coupled to the Higgs field is about 100. Provided that it is sufficient to model the fermions and gauge fields by a set of massless scalars, our results show that the LO thermal mass gives a good description of the blocking mechanism, conservatively within a factor of 2 in couplings and/or temperature.

Our results are applicable to the thermal blocking of curvaton preheating \cite{dani}. In this model, the thermal background is generated by the decay of the inflaton, before the start of the curvaton resonance. Our results confirm previous analytical results,  which showed that this thermal background produced substantially blocks curvaton preheating \cite{dani}. Without this blocking, the MCH model is not viable \cite{mch}. There may also be implications for preheating in the Higgs Inflation model \cite{HIpreheating} and for warm inflation \cite{Dymnikova:2001ga}.

A number of unresolved factors relating to the expansion of the Universe have not yet been included in our lattice simulations. These include the dilution of decay products, the thermalisation mechanism of the particles, the redshifting of modes in and out of bands and the decrease of the inflaton oscillation amplitude. In addition, there are outstanding issues relating to the field dynamics, including the back-reaction on the inflaton as a dynamical field, the decay of primary excitations into secondary fields, and the origin of the initial thermal background. All of these await further work, and details will have quantitative (but probably not qualitative) impact on the phenomenon of thermal blocking of preheating.

\vspace{0.2cm}

\noindent {\bf Acknowledgments:} RL is supported by the Alexander von Humboldt Foundation. AT is supported by a Young Investigator Grant from the Villum Foundation. Part of the simulations presented here were performed on the Abel machine under the Norwegian Supercomputing Network Notur.

\end{document}